\documentclass[prl,twocolumn,showpacs,preprintnumbers,amsmath,amssymb,superscriptaddress]{revtex4}
\usepackage{graphicx}% Include figure files
\usepackage{dcolumn}% Align table columns on decimal point
\usepackage{bm}% bold math

\begin{document}

%\preprint{APS/123-QED}

\title{Sensing of Fluctuating Nanoscale Magnetic Fields Using NV Centres in Diamond}
\author{Liam T. Hall}
 \email{lthall@physics.unimelb.edu.au}
\affiliation{Centre for Quantum Computing Technology, School of Physics, University of Melbourne, Victoria 3010, Australia}%
\author{Jared H. Cole}
\affiliation{Centre for Quantum Computing Technology, School of Physics, University of Melbourne, Victoria 3010, Australia}%
\affiliation{Institute f\"ur Theoretische Festk\"orperphysik and DFG-Centre for Functional Nanostructures (CFN), Universit\"at Karlsruhe, 76128 Karlsruhe, Germany}%
%\affiliation{Institute fur Theoretische Festkorperphysik and DFG-Centre for Functional Nanostructures (CFN), Universitat Karlsruhe, 76128 Karlsruhe, Germany}%
\author{Charles D. Hill}
\affiliation{Centre for Quantum Computing Technology, School of Physics, University of Melbourne, Victoria 3010, Australia}%
\author{Lloyd C.L. Hollenberg}
\affiliation{Centre for Quantum Computing Technology, School of Physics, University of Melbourne, Victoria 3010, Australia}%

%\affiliation{


\begin{abstract}

New magnetometry techniques based on Nitrogen-Vacancy (NV) defects in diamond allow for the imaging of static (DC) and oscillatory (AC) nanoscopic magnetic systems. However, these techniques require accurate knowledge and control of the sample dynamics, and are thus limited in their ability to image fields arising from rapidly fluctuating (FC) environments. We show here that FC fields place restrictions on the DC field sensitivity of an NV qubit magnetometer, and that by probing the dephasing rate of the qubit in a magnetic FC environment, we are able to measure fluctuation rates and RMS field strengths that would be otherwise inaccessible with the use of DC and AC magnetometry techniques. FC sensitivities are shown to be comparable to those of AC fields, whilst requiring no additional experimental overheads or control over the sample.
\end{abstract}

\pacs{03.65.Yz, 07.55.Ge, 07.79.-v}% PACS, the Physics and Astronomy
%taken from magnetometry papers
%07.55.Ge Magnetometers for magnetic field measurements
%07.57.Pt Submillimeter wave, microwave and radiowave spectrometers
%07.79.Pk Magnetic force microscopes
%others
%07.55.-w Magnetic instruments and components
%07.79.-v Scanning probe microscopes and components
\maketitle

%\section{\label{sec:level1}Introduction}

\begin{figure}
\includegraphics[width=2.8in,height=2.1in]{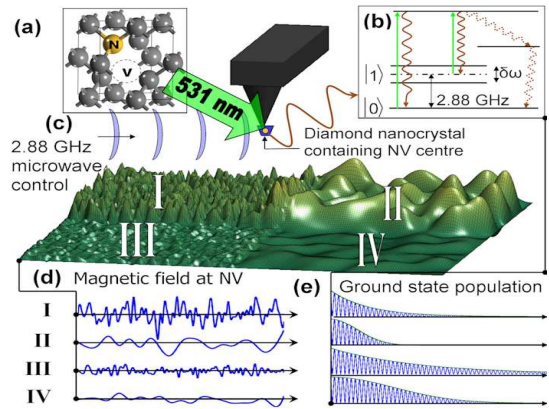}% Here is how to import EPS art
\caption{\label{scan}(colour online). Schematic of a scanning NV qubit magnetometer/decoherence probe for the detection of nanoscale field fluctuations. (a) NV-centre diamond lattice defect. (b) NV spin detection through optical excitation and emission cycle. (c) Microwave control of the spin state of the NV centre and 531 nm optical pulse for read-out. (d) Magnetic field signals $B(t)$ at the NV probe corresponding to regions I-IV of an inhomogeneous test sample with different fluctuation amplitudes and frequency spectra. (e) The corresponding qubit excited state populations $P(t)$ show that the regions can be distinguished by the dephasing information:  I: Strong, rapid fluctuations $\rightarrow$ fast exponential dephasing. II: Strong, slow fluctuations $\rightarrow$ fast Gaussian dephasing. III: Weak, rapid fluctuations $\rightarrow$ slow exponential dephasing. IV: Weak, slow fluctuations $\rightarrow$ slow Gaussian dephasing.
}
\end{figure}

The exploitation of controlled quantum systems as ultra-sensitive nanoscale detectors has tremendous potential to advance our understanding of complex processes occurring in biological and condensed-matter systems at molecular and atomic scales \cite{Has00,Kir95,Bla93}.
The stringent requirements for high sensitivity and spatial resolution has led to suggestions of using spin-based quantum systems as nanoscale magnetometers \cite{Che04}, or of imaging through detection of sample induced decoherence \cite{Col08}. One particularly attractive physical platform to implement these ideas is the Nitrogen-Vacancy (NV) centre in diamond [Fig.\,\ref{scan}(a)], chosen for its long coherence times at room temperature and convenient optical readout of the spin state \cite{Jel06} [Fig.\,\ref{scan}(b)]. As such, NV centres have been the focus of recent proposals to image static (DC) and oscillating (AC) magnetic fields \cite{Deg08,Tay08}, which have since been demonstrated experimentally \cite{Bal08,Bal09,Maz08}.

However, many important biological and condensed matter systems exhibit non-sinusoidal fluctuating magnetic fields with extremely low or zero mean values [Fig.\,\ref{scan}(d)]. An important question is therefore to what extent these quantum based magnetometry techniques are applicable to such situations. In this paper we address this by quantifying the detection sensitivities for these modes for samples with fluctuations characterized by the RMS field and dominant spectral frequency. The results indicate that by probing the dephasing rate of a spin qubit placed in such environments one can characterize the underlying fluctuation rates and RMS field strengths that would be otherwise inaccessible with the use of DC and AC magnetometry techniques, thereby opening the way for non-invasive nanoscale imaging of a range of biological and condensed matter systems.

\begin{figure}
\includegraphics[width=1.7in]{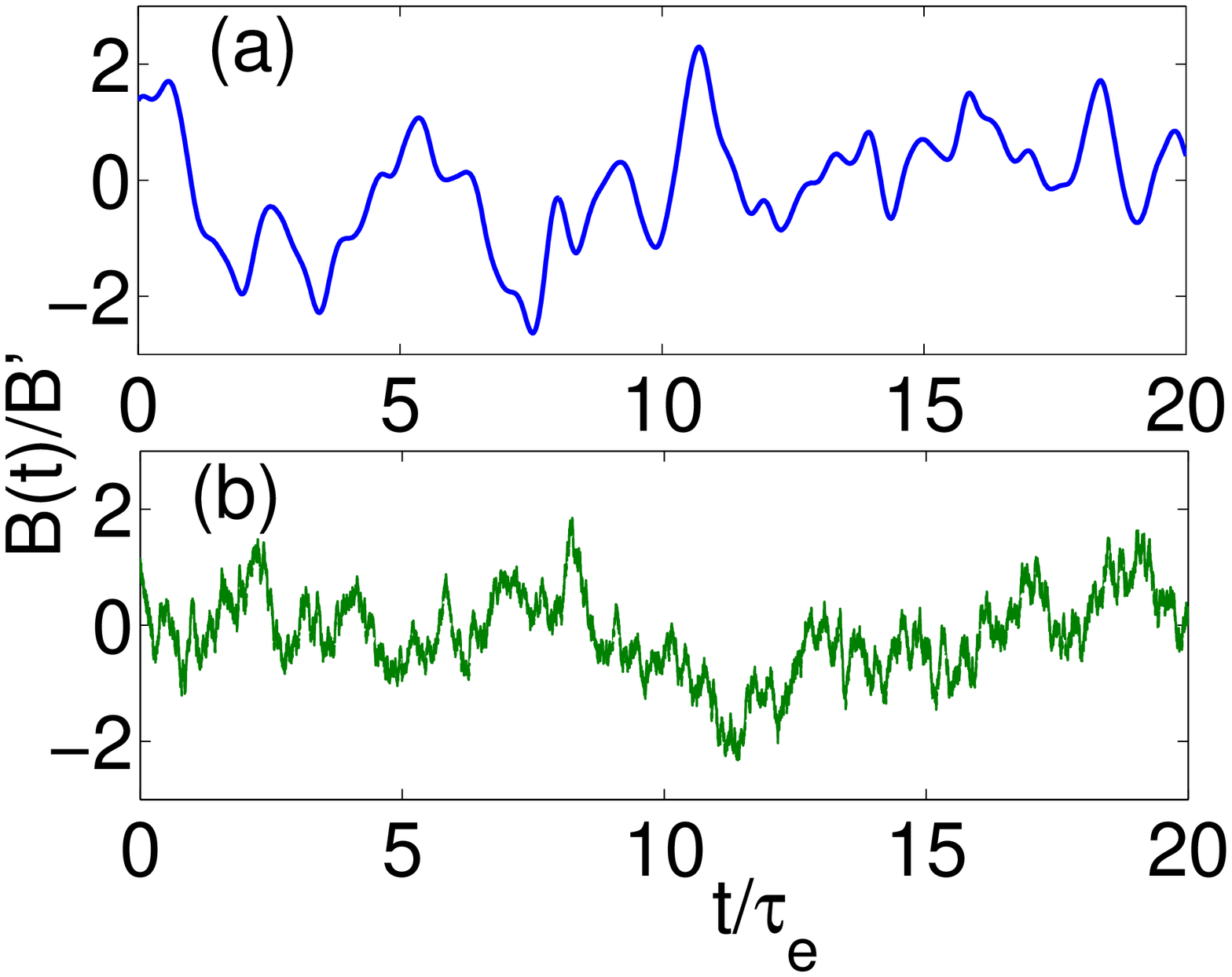}\includegraphics[width=1.7in,height=1.35in]{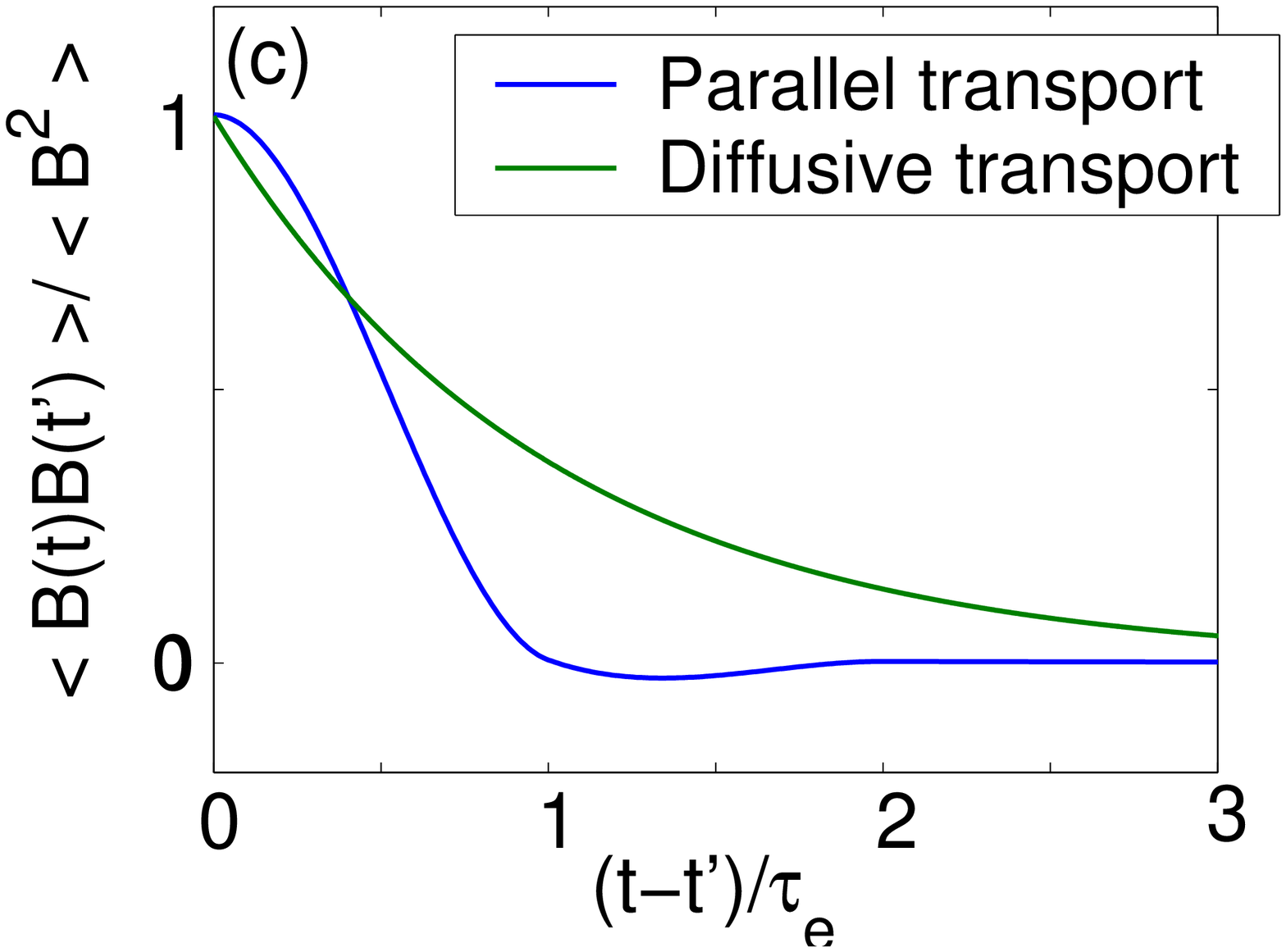}% Here is how to import EPS art
\caption{\label{BBcor}(colour online).  Typical magnetic field, $B(t)/B'$, for (a) a channel of dipoles in unidirectional motion, and (b) a self diffusing dipole bath. (c) Temporal correlation function $\langle B(t)B(t')\rangle/\langle B^2\rangle$ . Time axes %in (a), (b) \& (c)
are rescaled by $\tau_e$.}
\end{figure}

The theory behind the detection of magnetic fields using quantum systems is heavily reliant on the phase estimation program of quantum metrology, particularly the determination of coupling parameters that are constant in time. In the context of DC magnetometry, this corresponds to measurement of the first moment (the mean) of the magnetic field strength. For zero mean fields, complex microwave control pulse sequences are necessary. For fields exhibiting oscillatory (AC) time dependence with which either a spin-echo or Carr-Purcell-Meiboom-Gill \cite{Mei58} sequence may be synchronised, sensitivities are predicted to be as low as 3 nT$\mathrm{Hz}^{-1/2}$ \cite{Tay08}, based on the standard quantum limit. Excellent agreement between theory and experiment has been demonstrated in \cite{Maz08}. Such techniques require accurate knowledge of the field dynamics which may not be available, or more commonly, the field strength may exhibit a stochastic time dependence. Examples include nuclear dipole fields of ion channels \cite{Hil01} [Fig.\,\ref{BBcor}(a)] and lipid bi-layers in biological cell membranes \cite{Pat03}, Overhauser fields in Ga-As quantum dots \cite{Rei08}, and even self-diffusing water molecules \cite{Tik02,Rah79} [Fig.\,\ref{BBcor}(b)]. In what follows, we investigate the effects of a more general fluctuating (FC) field on the dephasing of a spin qubit as the primary detection mechanism, and the implications for the characterisation of the magnetic field from the surrounding environment.  In this sense, we are estimating the second moment of the environmental field strength, and the corresponding temporal dynamics.

A spin qubit placed in a randomly fluctuating magnetic environment will
experience a complex sequence of phase kicks, leading to an eventual dephasing of the population spectrum. For an NV centre, this will be in addition to the intrinsic sources of dephasing, which are due to paramagnetic impurities in the diamond lattice \cite{Chi06}. The dephasing rate can be quantified via repeated projective measurements of the qubit state, and the corresponding dephasing envelope, $\mathcal{D}(\tau)$,
can be determined via a suitably chosen quantum state reconstruction technique. We use the technique of Hamiltonian characterisation \cite{col05} rather than quantum tomography techniques, as it requires only a single measurement basis yet is robust in the presence of dephasing \cite{col06}.

The motivation for the environment model used here comes from consideration of magnetic dipoles in motion. Other models in which a two level system is coupled to a bath of bistable fluctuators have been previously considered \cite{Shn05,Sch06,Pal02,Gal06,Gut05,Mot06}. These models, however, do not capture the dephasing effects due to gradual transitions between environmental states in slowly fluctuating fields. Later we will show this to be of particular importance in the case of spin-echo based experiments. Additionally, these models require a large number of fluctuators to model a continuous signal. In contrast, we wish to consider the dephasing effects of small numbers of spins in motion.

Consider a qubit with gyromagnetic ratio $\gamma_p$ undergoing a $\frac{\pi}{2}\,-\,\tau\,-\,\frac{\pi}{2}$ Ramsey sequence \cite{Ebw90} in the presence of a classically fluctuating magnetic field, $B(t)$. An example of a fluctuating magnetic field due to a uni-directional spin current
[Fig.\,\ref{BBcor}(a)], and that of a bath of self-diffusing spins [Fig.\,\ref{BBcor}(b)]. The field has mean $\langle B \rangle \equiv B_0$, standard deviation $\sqrt{\langle B^2 \rangle - \langle B \rangle ^2}\equiv B'$, and typical fluctuation rate $f_e\equiv1/\tau_e$, where $\tau_e$ is the characteristic correlation time of the external field [Fig.\,\ref{BBcor}(c)]. %Such fields arise, for example, when the qubit is brought near a substrate of classical magnetic dipoles whose translational motion produces a fluctuating classical magnetic field.
This gives rise to two natural frequency scales, given by $\omega_0 = \gamma_pB_0$ and $\omega' = \gamma_pB'$.  The average precession frequency of the qubit is set by $\omega_0$, and is found to be decoupled from all dephasing effects for cases where $\omega',f_e\ll\omega_0$. Additional relaxation processes may dominate the qubit evolution when this condition is violated, however such cases are not considered here since we are interested in the characterisation of weak magnetic fields. The nature of the dephasing felt by the qubit will depend on the fluctuation rate of the environment, $f_e$, or more specifically the magnitude of the quantity defined by $\Theta \equiv  f_e/\omega'$.
In the case of $\Theta\gg1$, or fast-fluctuation limit (FFL), the qubit will experience many environmental fluctuations during its natural timescale. Whilst $B'$ need not necessarily be normally distributed, the accumulated phase error of the qubit at some time $t\gg1/f_e$ will be normally distributed by way of the Central Limit Theorem. As such, the variance of the phase error at time $t\gg1/\omega'$ will be $\langle \Delta \phi ^2\rangle \sim t\gamma_p^2B'^2/f_e$, giving rise to an FFL dephasing rate of
\begin{eqnarray}
\Gamma_{\mathrm{fast}}\left(B',f_e\right) &=& \frac{\gamma_p^2B'^2}{2f_e}\label{fastdeph}.
\end{eqnarray} This is akin to the motional narrowing result from NMR \cite{Kit} and reproduces the ubiquitous exponential dephasing envelope %(ie that which modulates the free qubit evolution)
given by $\mathcal{D}_{\mathrm{fast}}(t) = \exp\left(-\Gamma_{\mathrm{fast}}t\right)$.

In the slow-fluctuation limit (SFL), where $\Theta\ll1$, we note that the magnetic field %will vary slowly, and thus
may be locally approximated by a Taylor expansion in $t$ about some initial time $t_0$: $B(t) = \sum_{k=0}^N\left.\frac{1}{k!}\frac{d^kB}{dt^k}\right|_{t_0}\left(t-t_0\right)^k \equiv \sum_{k=0}^Na_k\left(t-t_0\right)^k$, where each of the $a_k$ has a specific statistical distribution containing information about the $k$th order derivative of $B(t)$, and thus gives rise to a different dephasing channel.

For the special case where the $a_k$ are normally distributed with mean $\mu_k$ and variance $\sigma_k^2$ (as consistent with random dipole motion), the resulting density matrix following the free evolution time $\tau$, but prior to the second $\pi/2$ pulse is defined by $\rho_{11} = \rho_{22} = 1/2$, and $\rho_{12} = \rho_{21}^* = \prod_{k=0}^\infty\mathcal{D}_{\mathrm{slow}}^{(k)}(\tau)\Omega_{\mathrm{slow}}^{(k)}(\tau)$; where
\begin{eqnarray}
\mathcal{D}_{\mathrm{slow}}^{(k)}(t) &=& \exp\left[-\left(\Gamma_\mathrm{slow}^{(k)}t\right)^{2k+2}\right],\,\,\,\mathrm{and}\label{hierachy}\\
\Omega_{\mathrm{slow}}^{(k)}(t) &=& \exp\left[-i\left(\omega_\mathrm{slow}^{(k)}t\right)^{k+1}\right].
\end{eqnarray}
Thus we see the emergence of a hierarchy of dephasing and beating channels, with the dephasing rates and beat frequencies of the $k^{\mathrm{th}}$ channel given by
\begin{eqnarray}
  \Gamma_\mathrm{slow}^{(k)} &=&  \left(\frac{1}{\sqrt2}\frac{\sigma_k\gamma_p}{k+1}\right)^{1/(k+1)}\label{slowdeph}\\
  \omega_\mathrm{slow}^{(k)} &=& \left(\frac{\mu_k\gamma_p}{k+1}\right)^{1/(k+1)}
\end{eqnarray} respectively. In the case of the zeroth order channel this corresponds to the rigid lattice result from NMR \cite{Kit}, and we have $\sigma_0^2 = \langle B^2\rangle-\langle B\rangle^2$. This effect will be suppressed by a spin echo pulse sequence. For the first order channel, we may approximate $\sigma_1^2 \sim \left(\langle B^2\rangle -\langle B\rangle^2\right)f_e^2$.

The relative contributions of each channel to the overall dephasing rate of the qubit depend explicitly on the dynamics of the field, however, it should be noted that dominance of the zeroth order channel (ie $\Gamma_\mathrm{slow}^{(0)} > \Gamma_\mathrm{slow}^{(j)},\,\,\forall\,\,j\geq1$) is a necessary and sufficient condition for the system to exist in the slow fluctuation regime, $\Theta\ll1$. This justifies the use of the Taylor expansion, since the resulting polynomial may be well approximated by a low-order truncation.

\begin{figure}
\includegraphics[width=1.7in]{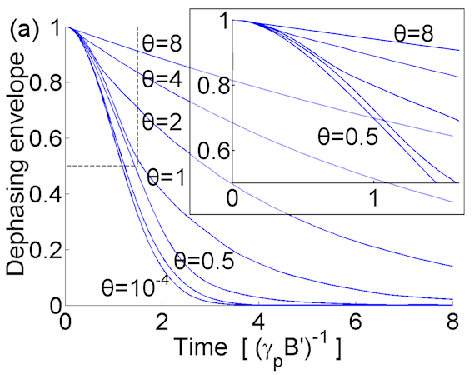}\includegraphics[width=1.7in,height=1.35in]{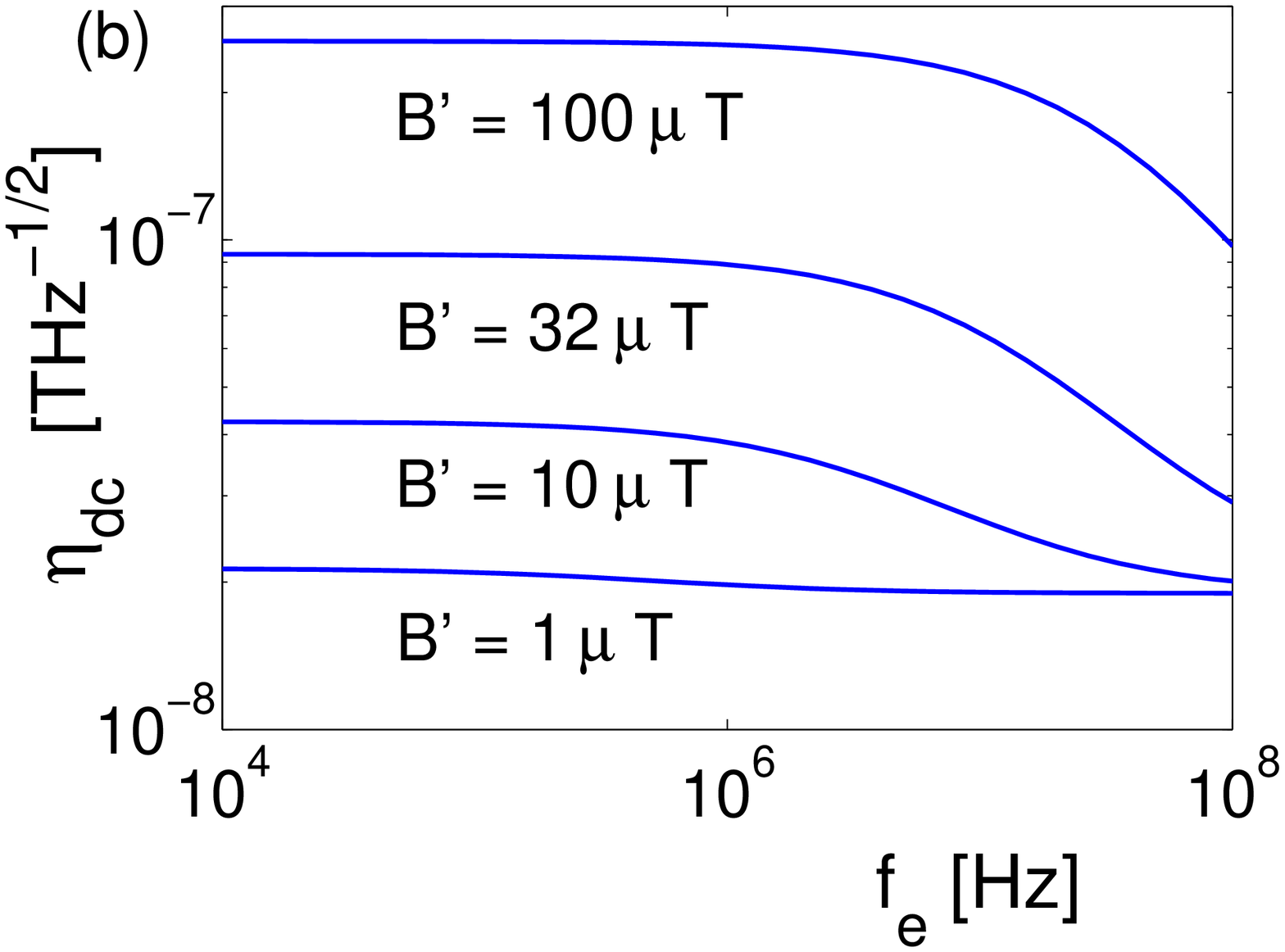}% Here is how to import EPS art
\caption{\label{Ds}(colour online). (a) Plot of simulated dephasing envelopes for $N_s=10^4$ runs, showing agreement with Eqs. \ref{fastdeph} \& \ref{hierachy}. Time is in units of $\left(\gamma_pB'\right)^{-1}$. (insert) Zoomed plot showing that fast fluctuating environments still exhibit non-exponential dephasing for short timescales $\tau:\omega'\tau<\sqrt2/\Theta$. (b) log-log plot of DC magnetic field sensitivity, $\eta_{\mathrm{dc}}$, as a function of $f_e$ for different contours of $B'$. Assumed parameter values are $T_2^* = 1\,\mu\mathrm{s},\,\mathrm{and}\,C = 0.3$. }
\end{figure}

The intermediate regime of $\Theta \sim 1$ is more complicated. Fig.\,\ref{Ds}(a) shows dephasing envelopes for various values of $\Theta$. For times much longer that $\tau_e$, pure exponential dephasing behaviour is observed in all cases (with dephasing rate $\Gamma_\mathrm{fast}$), however fast fluctuating environments still exhibit slow (Gaussian) dephasing behaviour on timescales $\tau$ where $\omega'\tau < \sqrt2/\Theta$. If $\Theta$ is large, contributions to $\mathcal{D}$ from the $\Gamma_\mathrm{slow}^{(k)}$ will decay rapidly. The abrupt transition from $\mathcal{D}_\mathrm{slow} \rightarrow \mathcal{D}_\mathrm{fast}$ is shown more clearly in the corresponding insert.

For the purpose of comparison with existing spin-based magnetometer proposals, we take the NV centre as our example qubit. The Hamiltonian used to describe the time evolution of an NV-centre is given by $\mathcal{H} = \mathbf{S}\cdot\mathbf{D}\cdot\mathbf{S} + \hbar\gamma_p\mathbf{B}\cdot\mathbf{S} +\mathcal{H}_\mathrm{other}$, where $\mathcal{H}_\mathrm{other}$ describes higher order effects such as hyper-fine splitting, interaction with optical fields, etc.\  which can be ignored in the present context. We consider weak external fields such that $\mathcal{O}\left(\hbar\gamma_p\mathbf{B}\cdot\mathbf{S}\right)\ll \mathcal{O}\left(\mathbf{S}\cdot\mathbf{D}\cdot\mathbf{S}\right)$, thereby ensuring the crystal-field splitting tensor, $\mathbf{D}$, sets the quantisation axis of the NV centre, and that $\omega'\ll\omega_0$ (even in the FFL).

The shot-noise-limited DC magnetic field sensitivity for an NV-based magnetometer subject to a Ramsey-style pulse sequence is given by \cite{Tay08}
\begin{eqnarray}
  \eta_{\mathrm{dc}} \equiv B_{\mathrm{min}}\sqrt T \approx \frac{1}{\gamma_p C \sqrt {\tau}}\label{dcsen},
\end{eqnarray} where $\sqrt T$ and $C$ represent the combined effects of spin projection and photon shot noise for $N_s$ measurements ($C\rightarrow 1$ for the ideal case), $\tau$ is the free evolution time of the qubit in a given experiment, and $T = N_s\tau$ is the total averaging time for $N_s$ such experiments. Dephasing times due to the interaction of the NV centre with nearby paramagnetic lattice impurities will in general be different for different centres and will thus require individual characterisation. For comparison with \cite{Tay08}, we use the commonly accepted value of $\tau = T_2^*\sim 1\,\mu$s.

We emphasise here that Eqn. \ref{dcsen} applies solely to the imaging of DC magnetic fields where the dephasing of the qubit is exclusively due to intrinsic crystal effects. If the sample being imaged produces a fluctuating field of sufficient amplitude, the dephasing time ($1/\Gamma$) may be shorter than $T_2^*$, resulting in poorer static field sensitivity. In this context, $\eta_{\mathrm{dc}}$ refers to the sensitivity with which the mean field, $\langle B\rangle$, may be measured as the field fluctuates over the course of the experiment.
To gain insight into the effect of fluctuating magnetic fields on the DC field sensitivity, we consider again a $\frac{\pi}{2}-\tau-\frac{\pi}{2}$ sequence. The DC sensitivity as a function of $B'$ and $f_e$ is shown in Fig.\,\ref{Ds} (b). From this, we see that fluctuating environments can have a dramatic effect on the DC field sensitivity of an NV based magnetometer, depending on both field strength and fluctuation frequency.

\begin{figure}
\includegraphics[width=1.7in]{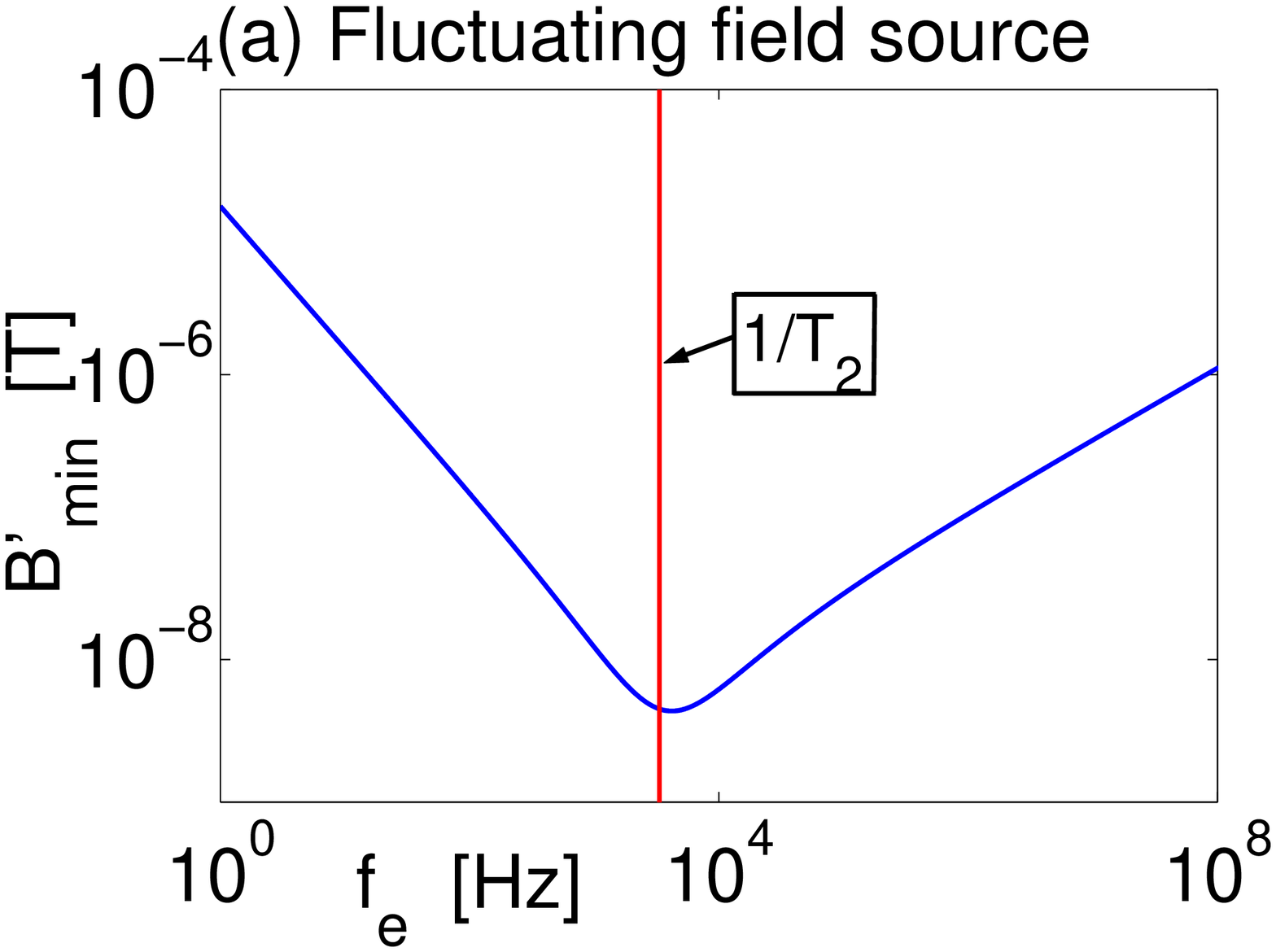}\includegraphics[width=1.7in]{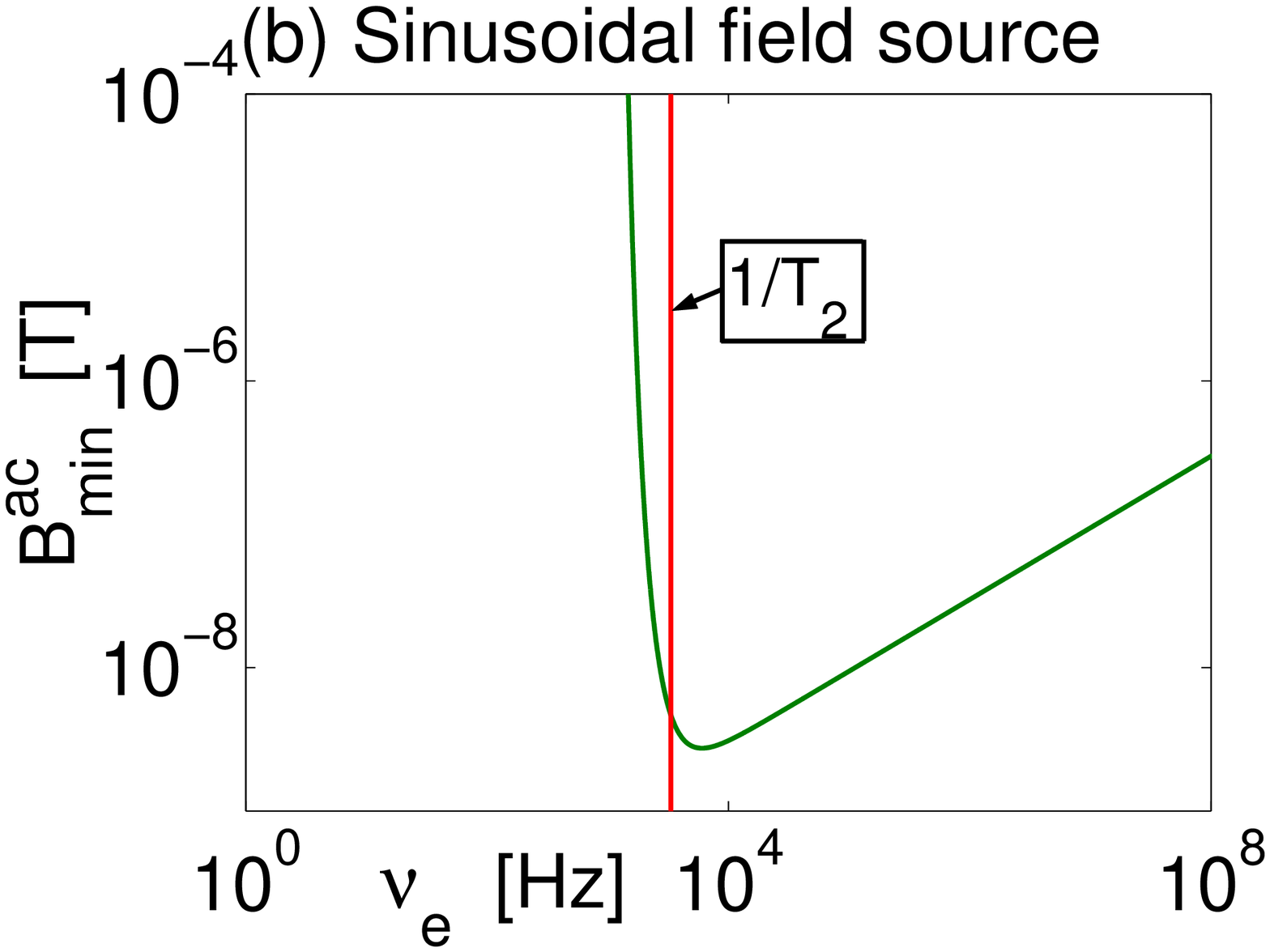}% Here is how to import EPS art
\caption{\label{sens}(colour online). (a) Minimum resolvable FC field strength, $B_{\mathrm{min}}$, vs environmental fluctuation rate, $f_e$, for $T = 1$ s averaging time. In contrast to the AC case, an FC detection requires no prior knowledge of fluctuation timescales. (b) Minimum resolvable AC field amplitude, $B_{\mathrm{min}}^{\mathrm{ac}}$, vs field oscillation frequency, $\nu_e$, for $T = 1$ s averaging time. Here we have assumed that the AC field is initialised in phase with the probe qubit, and that the oscillation frequency is known accurately enough that the $\pi$ pulse of a spin-echo sequence coincides with the first zero-crossing of the field. This will become increasingly difficult with increasing $\nu_e$, leading to additional sources of dephasing.}
\end{figure}

We now turn our attention to the the sensing of magnetic field fluctuations themselves, including the case of zero-mean magnetic fields. Using coherent control techniques (spin-echo for example), we may extend the dephasing time of the NV-centre to $T_2\sim 300\,\mu$s, as dictated by the 1.1\% carbon-13 content in the lattice. The case of perfectly oscillatory magnetic fields has been considered in detail in \cite{Tay08}, in which AC sensitivities may be as low as 3 nT $\mathrm{Hz}^{-1/2}$ [Fig.\,\ref{sens}(b)]. This technique requires the $\pi$ pulse to coincide with the first zero-crossing of the magnetic field, which requires an accurate knowledge of the oscillation frequency. This may prove difficult unless the frequency is externally controlled. Furthermore, accurate control will become difficult at high frequencies, and this may lead to further dephasing.

Rather than considering an AC field, we now study the magnetometer's sensitivity to a more general fluctuating field via consideration of the induced dephasing rate \cite{Col08}. For a $\frac{\pi}{2}-\frac{\tau}{2}-\pi-\frac{\tau}{2}-\frac{\pi}{2}$ pulse sequence, the probe will show decreased sensitivity to environments exhibiting fluctuation frequencies, $f_e$ less than $1/\tau$. For fast fluctuating fields, the effect will be negligible. For the imaging of slowly fluctuating fields, this may appear problematic, however complete insensitivity only comes with $f_e\rightarrow0$. A spin echo sequence will modify the $\mathcal{D}_\mathrm{slow}^{(k)}$ via $\Gamma_\mathrm{slow}^{(k)} \longmapsto \left(1-2^{-k}\right)\Gamma_\mathrm{slow}^{(k)}$, thus only the effects of the zeroth order dephasing channel will vanish.

Perturbations on the dephasing rate may be measured from
$\left(1-\mathcal{D}\right)_{\mathrm{min}}=\frac{\exp\left[\left({\tau}/{T_2}\right)^n\right]}{C\sqrt{N_s}}$,  where $n$ describes the shape of the spin-echo dephasing envelope as dictated by the presence of carbon-13 nuclei in the lattice, which for present purposes may be taken as $n=3$ \cite{Chi06}. This implies an optimal free-evolution time of $\tau \sim T_2/\sqrt[3]{6}$.

Thus we find that perturbations on the $1/T_2$ dephasing rate as slow as 200 Hz for exponential dephasing and 800 Hz for Gaussian dephasing may be detected by this method after 1 s of averaging time. By performing measurements of the total dephasing rate, $\Gamma$, both the field variance and average fluctuation rate may be inferred from equations \ref{fastdeph} \& \ref{slowdeph}. Of course, the question remains of which fluctuation regime a given sample system resides in. In the absence of any prior knowledge of the environment being measured, this question may be answered via determination of the shape of the dephasing envelope, a task to which the Hamiltonian Characterisation method is well suited \cite{Col08}.

The optimal fluctuating magnetic field sensitivity will occur when the $\Theta\sim1$ condition is satisfied, since this ensures maximal dephasing for a given field variance. Considering the special case of pure exponential decay, we therefore expect an optimal sensitivity of $\eta_{\mathrm{fc}} = \frac{e^{1/6}}{C\gamma_p\sqrt{T_2}}$, which for $C=0.3$ gives $\eta_{\mathrm{fc}} = 1.7\,\mathrm{nT}\mathrm{Hz}^{-1/2}$.
In practice however, such sensitivity may be difficult to realise due to memory effects in the fluctuating environment. For systems that satisfy $\Theta \gg 1$, thus exhibiting long-time exponential dephasing behaviour, Gaussian dephasing is still exhibited for $\tau < 1/f_e$ [Fig.\,\ref{Ds}(a)]. For spin-echo experiments, the effect is worsened as the dominant contribution to $\mathcal{D}_\mathrm{slow}$ comes from $k=1$. Taking this into consideration, the minimum resolvable field obtained after $T = 1$ s averaging time is plotted in Fig.\,\ref{sens}(a) as a function of environment fluctuation frequency. We see that fluctuating field strengths as low as 4.5 nT may be achievable after $T = 1$ s averaging time, or some $N_s \sim 3000$ measurements. It is also evident that the qubit will be sensitive to FC fields fluctuating on timescales much slower than $1/T_2$. This is in direct contrast with the AC case, which shows poor sensitivity to fields oscillating with periods less than $T_2$ [Fig.\,\ref{sens}(b)].

We have theoretically investigated the effects of a fluctuating magnetic field on an NV centre spin qubit. This analysis was used to place new limits on the sensitivity with which the mean field strength may be measured. Furthermore, we have built upon the idea of decoherence microscopy \cite{Col08} to theoretically demonstrate the ability of an NV centre to measure field strengths and fluctuation rates of randomly fluctuating magnetic fields. This analysis shows that the methods presented here require no experimental resources beyond those of existing techniques, no prior control or knowledge of the external field, and thus may be implemented with current technology.

We gratefully acknowledge discussions of the subject matter with A. Greentree, M. Testolin, F. Jelezko and J. Wrachtrup. This work was supported by the Australian Research Council (ARC) and the Alexander von Humboldt Foundation.

\bibliography{sensbib}% Produces the bibliography via BibTeX.

\end{document}